\documentclass[iop]{emulateapj}
\usepackage{ifpdf} 
\pdfoutput=1  % arxiv.org needs this in the first 5 
\usepackage{graphicx}
 \usepackage[usenames, dvipsnames]{xcolor}
\definecolor{lightgreen}{HTML}{228B22}
\definecolor{darkgreen}{HTML}{006400}
\usepackage{cmap} 
\usepackage[colorlinks, bookmarks, breaklinks, pdftitle={HETMGS data release},pdfauthor={Remco C. E. van den Bosch}]{hyperref}
\hypersetup{linkcolor=black,citecolor=black,filecolor=black,urlcolor=darkgreen,pdfborderstyle={/S/U}}   
\usepackage[english]{babel}
\usepackage[activate=true,kerning=true,tracking=true,spacing=true,babel,final]{microtype}  % pdfsync does not work with 
\usepackage[yyyymmdd,hhmmss]{datetime}
\usepackage{afterpage}
\RequirePackage[normalem]{ulem} 
   
\usepackage{mathptmx}
\usepackage[T1]{fontenc}

\def\kms{km\,s$^{-1}$}
\def\dgr{$^\circ$}
\def\Mbh{$M_\bullet$}
\def\Msun{M$_\odot$}

\def\farcs{\hbox{$.\!\!^{\prime\prime}$}}
\def\arcs{\hbox{$^{\prime\prime}$}}
\def\arcm{\hbox{$^{\prime}$}}
\def\Msigma{\mbox{\Mbh--\,\relsize{-1}$\sigma$\relsize{+1}}}
\def\MsigmaE{\mbox{\Mbh--\,\relsize{-1}$\sigma_e$\relsize{+1}}}
\def\MsigmaC{\mbox{\Mbh--\,\relsize{-1}$\sigma_c$\relsize{+1}}}
\def\Mbulge{\mbox{\Mbh--M$_{{\mathrm{bul}}}$}}
\def\MLrel{\mbox{\Mbh--L$_{{\mathrm{bul}}}$}}
\def\MLtot{\mbox{\Mbh--L$_{{\mathrm{tot}}}$}} 
\def\ML{\mbox{\Mbh--L}} 
\def\SOI{sphere of influence} 
\def\SOIs{spheres of influence} 

\def\SoI{\ensuremath{\theta_{{i}}}} 
\def\RoI{\ensuremath{r_{{i}}}} 
\def\Amu{\ensuremath{\langle \mu_e \rangle}}
\def\ATLAS{\mbox{ATLAS$^{\mathrm{{3D}}}$}} 
\def\ATLAScap{\mbox{ATLAS${\mathrm{{3D}}}$}} 
\def\HETMGS{\mbox{HETMGS}}
\def\CALIFA{\mbox{CALIFA}}

\newcommand{\referee}[1]{#1}
\newcommand{\DIFdel}[1]{}
\newcommand{\todo}[1]{} 
\newcommand{\raid}[1]{}
\def\websiteurl{http://www.mpia.de/~bosch/hetmgs}
\usepackage{relsize}
\usepackage{longtable}
\def\centralsigvssige{$\sigma_e=(18\pm5) + (0.85\pm0.03) \sigma_c$ with an intrinsic scatter of $8\pm6$}
\def\msigmacoef{($8.37\pm0.07$, $5.49\pm0.40$, $0.43\pm0.23$) and ($8.24\pm0.06$, $5.28\pm0.37$, $0.40\pm0.21$)}
\def\quarticsurfacecoef{$\log\sigma = 6.39  -0.32 \Amu + 0.004\Amu ^2  -0.10 \log R_e +2.14(\log R_e)^2 + 0.27\Amu \log R_e$}
\def\HETtmassN{23000} % number of objects in the 2mass selection  
\def\HETParentSample{29000} % Total number of objects in catalog 

\def\HETNGals{1022}
\def\HETSDSSoverlap{148}
\def\HETSDSSstddev{11}
\def\HETrelstddevOfSdsssubset{35}
\def\HETSDSSVELstddev{20}
\def\HETSDSSVELmean{23}
\def\SDSSvelformerror{3}  % Formal error on the SDSS overlap helio velocities
\def\HETHLoverlap{471}
\def\SDSSHLoverlap{1677}
\def\HETSDSSHLoverlap{73}
\def\HETVELmean{-38}
\def\HETVELstddev{43}
\def\HETNagn{50}
\def\HETNwithoutdisp{471}
\def\HETNwithdisp{546}
\def\HETNwithbigsoi{114}
\def\HETNwithbigsoifromfp{31}
\def\SOIcompletefif{0.045} % HET SOI completeness: The survey is 50% complete down to this SOI
\def\SOIcomplete{0.065} % HET SOI completeness: The survey is 95% complete down to this SOI
\def\Ngalbigsoisig{100}
\def\Ngalbigsoisigfrac{17}

\def\HETNknownbh{53}
\def\Nknownbh{63}
\def\HETmeddist{65} % median distance of HETMGS

\def\fpdissel{$0.10$}

\shortauthors{Remco van den Bosch} 
\shorttitle{HET Massive Galaxy Survey}
\slugcomment{Accepted to ApJs.}

\begin{document}

\title{Hunting for Supermassive Black Holes in Nearby Galaxies with the Hobby-Eberly Telescope}

\author{Remco C. E. van den Bosch\altaffilmark{1,5}, Karl Gebhardt\altaffilmark{2}, \\ Kayhan G\"ultekin\altaffilmark{3}, Akin Y{\i}ld{\i}r{\i}m\altaffilmark{1} and Jonelle L. Walsh\altaffilmark{2,4}}

\altaffiltext{1}{Max-Planck Institut f\"ur Astronomie, K\"onigstuhl 17, D-69117 Heidelberg, Germany}
\altaffiltext{2}{Department of Astronomy, The University of Texas at Austin, 2515 Speedway, Stop C1400, Austin, TX 78712, USA}
\altaffiltext{3}{Department of Astronomy, University of Michigan, Ann Arbor, Michigan 48109, USA}
\altaffiltext{4}{George P. and Cynthia Woods Mitchell Institute for Fundamental Physics and Astronomy, and Department of Physics and Astronomy, Texas A\&M University, College Station, TX 77843, USA}
\altaffiltext{5}{email: bosch@mpia.de}

 \begin{abstract}

   We have conducted an optical long-slit spectroscopic survey of \HETNGals\ galaxies using the 10m Hobby-Eberly Telescope (HET) at McDonald Observatory. The main goal of the HET Massive Galaxy Survey (HETMGS) is to find nearby galaxies that are suitable for black hole mass measurements. In order to measure accurately the black hole mass, one should kinematically resolve the region where the black hole dominates the gravitational potential. For most galaxies, this region is much less than an arcsecond. Thus, black hole masses are best measured in nearby galaxies with telescopes that obtain high-spatial resolution. The HETMGS focuses on those galaxies predicted to have the largest sphere-of-influence, based on published stellar velocity dispersions or the galaxy fundamental plane. To ensure coverage over galaxy types, the survey targets those galaxies across a face-on projection of the fundamental plane. We present the sample selection and resulting data products from the long-slit observations, including central stellar kinematics and emission line ratios. The full dataset, including spectra and resolved kinematics, is available online. Additionally, we show that the current crop of black hole masses are highly biased towards dense galaxies and that especially large disks and low dispersion galaxies are under-represented. This survey provides the necessary groundwork for future systematic black hole mass measurement campaigns. \\ Data, including spectra, are available at \url{\websiteurl}.
    
\end{abstract}


\newcommand{\plothetvscagcvshl}{
\begin{figure*}
\centering
\includegraphics[trim=0 0 0 0.9cm,clip=true,width=0.95\textwidth]{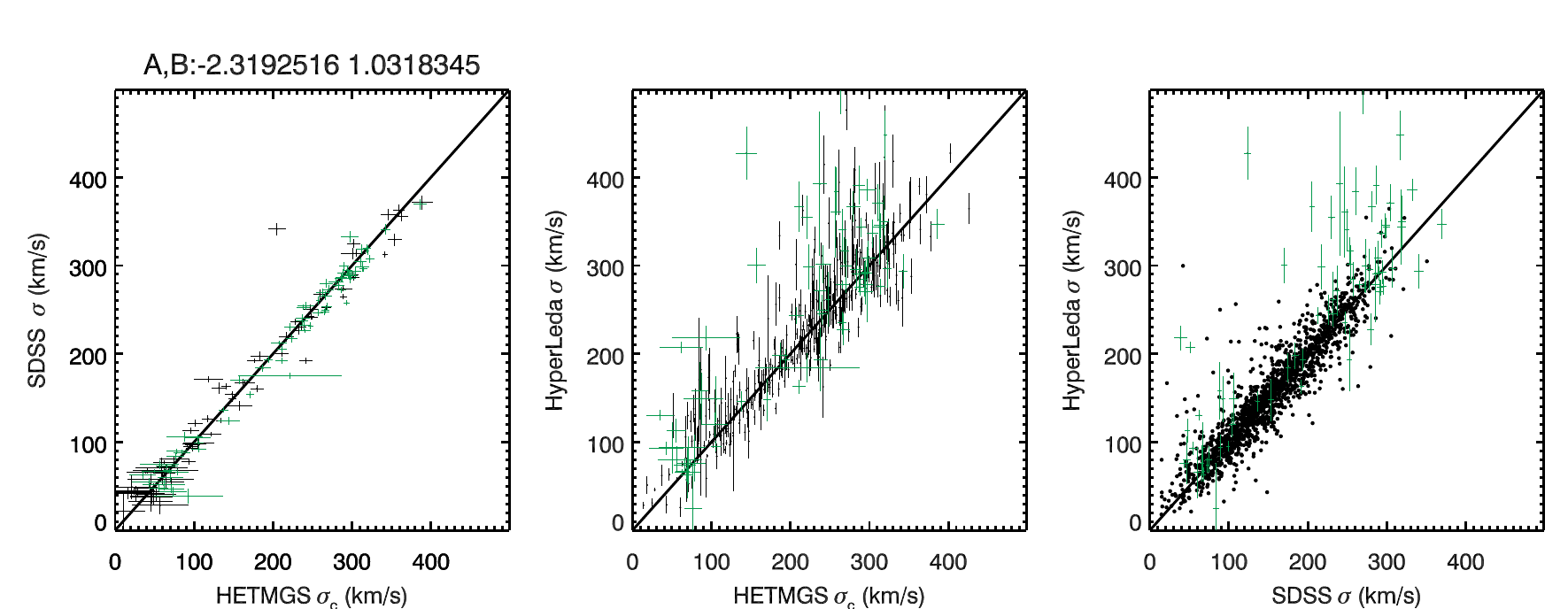}
\caption{A comparison of the HETMGS central velocity dispersion
  measurements to literature values from the SDSS and the
  HyperLeda database. The HETMGS values are measured in the central
  3.5$\times$2\arcsec\ or 3.5$\times$1\arcsec. The SDSS measurements
  are from 3\arcsec\ fibers, while the HyperLeda database draws from
  the amalgamated literature. The \HETSDSSHLoverlap\ green objects
  represent the galaxies present in all three samples. The HETMGS
  dispersions agree well with SDSS. The HyperLeda dispersion are very
  unreliable, especially for values above 300 \kms. In the middle and right panel, the error bars on the $x$-axis and both axis are suppressed for clarity, resp.}
\label{fig:hetvsvagcvshl}  
\end{figure*}}

\newcommand{\plotHETfp}{
\begin{figure}
\centering
\includegraphics[width=0.8\columnwidth,trim=0cm 0cm 0 8.0cm,clip=true]{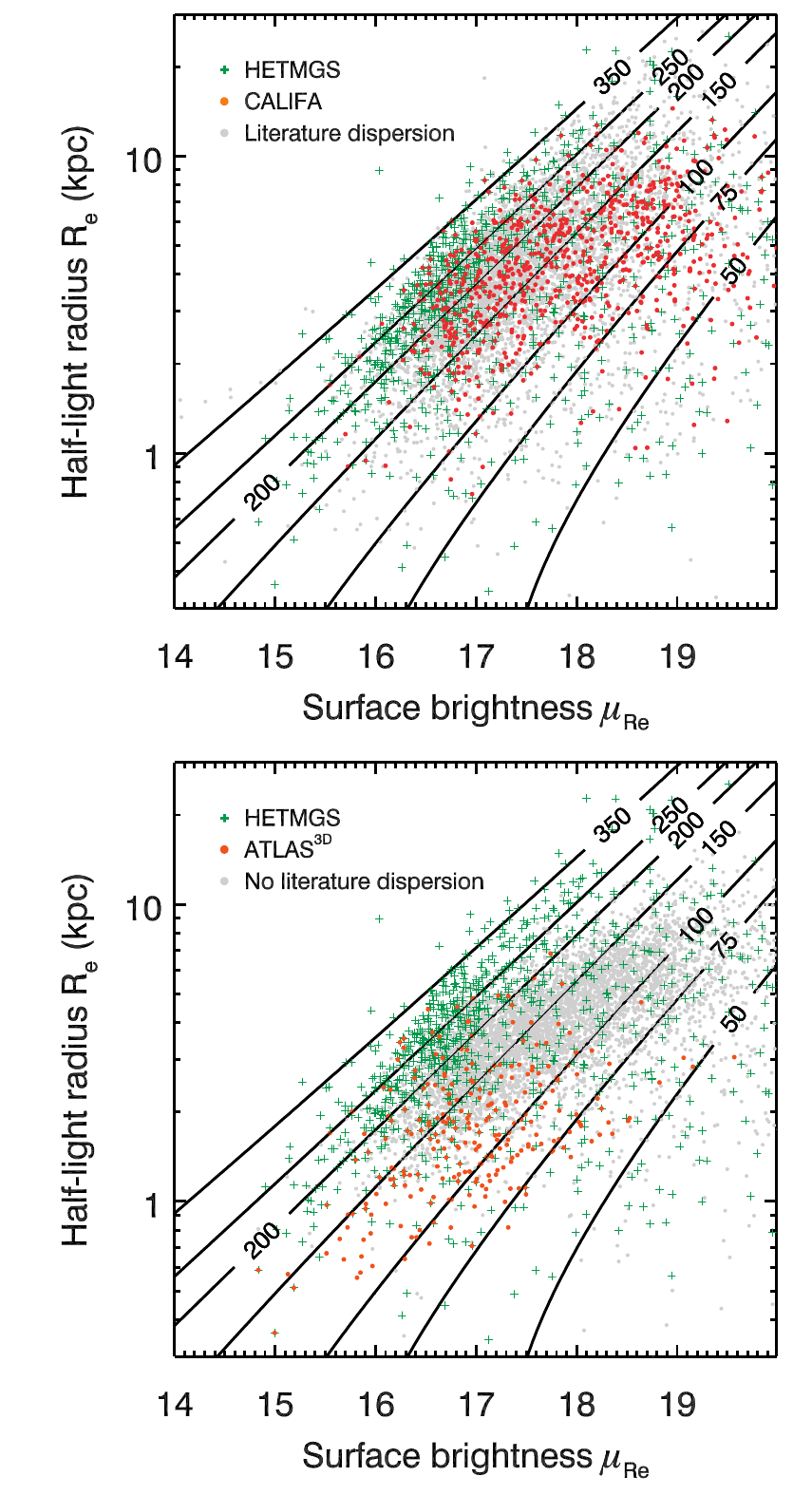}  

\caption{
The distribution of average surface brightness \protect{$\Amu$} and
half light radius $R_e$ of surveyed galaxies. The \textbf{left} panel
shows the \HETMGS, \CALIFA\ mother sample and literature dispersions. The \textbf{right} panel
shows the \HETMGS, \ATLAScap\ and un-surveyed galaxies in the parent
sample. The \HETMGS\ samples different types of galaxies than both
\ATLAScap\ and \CALIFA. Faint galaxies in the \CALIFA\ mother sample are not
shown as they are too faint to be in the 2MASS XSC. The over-plotted
lines indicate the velocity dispersion predictions in \kms\ using our
calibration of the fundamental plane (\S\ref{sec:dispersion_estimates_from_fundamental_plane}).}
\label{fig:2masspf}
\end{figure}
}

\newcommand{\plotsigmapred}{
\begin{figure}
\centering
\includegraphics[width=0.95\columnwidth,trim=0 16.0cm 0.0cm 0.0cm,clip=true]{figures/HETfp}  
\caption{
Velocity dispersion predictions from the Fundamental Plane. The figure shows measured stellar velocity dispersion versus predicted velocity dispersion from the fundamental plane (\S\ref{sec:dispersion_estimates_from_fundamental_plane}). The comparison is done with galaxies that have a HETMGS dispersion (green crosses) and the grey dots are SDSS galaxies. The thick and dashed lines show equality and the 0.09 dex RMS scatter, respectively.}
\label{fig:sigmapred}
\end{figure}
} 
  
\newcommand{\plothetsois}{\begin{figure*}
\centering
\includegraphics[width=1.0\textwidth]{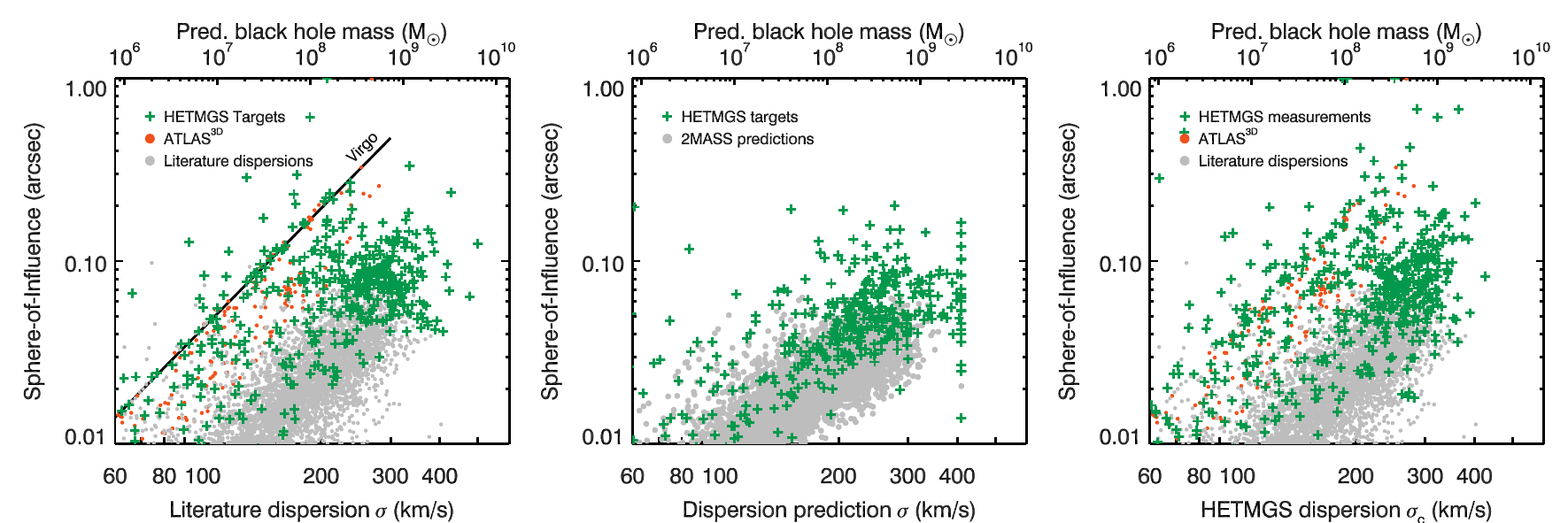}
\caption{The HETMGS targets are shown as function of velocity
  dispersion, estimated black hole mass from \Msigma, and estimated
  sphere of influence. The primary objective of the HETMGS is to
  observe galaxies with the highest predicted \SOIs . The survey
  succeeded in observing 95\% of all galaxies with \SoI\ greater than
  \SOIcomplete\arcs. \textbf{Left} panel: galaxies from the parent sample  with literature
  dispersions and their \SOIs\ predicted using \Msigma\ are shown as grey
  circles and the HETMGS targets that were observed from this subset
  are plotted as green crosses. The black line indicates the where galaxies at the distance of the Virgo cluster lie. \textbf{Middle} panel: the predicted
  velocity dispersions from the galaxy fundamental plane and the
  estimated \SoI\ of all 2MASS galaxies in the parent sample without
  literature dispersions are given by the grey circles, while the
  HETMGS targets selected from this subset are shown as green
  crosses. \textbf{Right} panel: all \HETNGals\ galaxies in the HETMGS
  sample are displayed with the green crosses. The velocity dispersion
  values shown are the measurements made from the HET data. The grey
  circles are the remaining parent sample galaxies with a literature dispersion. The
  red points are galaxies in the \ATLAScap\ survey, but not in
  HETMGS. For reference the \emph{HST} resolution is 0\farcs1
  FWHM. Galaxies with a known black hole masses or not visible from the HET are not
  shown.}
\label{fig:hetsois}
\end{figure*}}

\newcommand{\plotHETapparent}{\begin{figure}
\centering
\includegraphics[width=\columnwidth]{figures/HETapparent}
\caption{Apparent properties of the galaxies with large SOIs. The red circles show the literature black hole masses and the green crosses represent the HETMGS galaxies with SOIs$>$0.1\arcsec. Larger symbols indicate bigger (predicted) black hole masses. They grey dots represent the 2MASS parent sample. }
\label{fig:HETapparent}
\end{figure}}

\newcommand{\plotspectrum}{\begin{figure*}
\centering
\includegraphics[width=0.95\textwidth,trim=0 6cm 0 6cm ,clip=true]{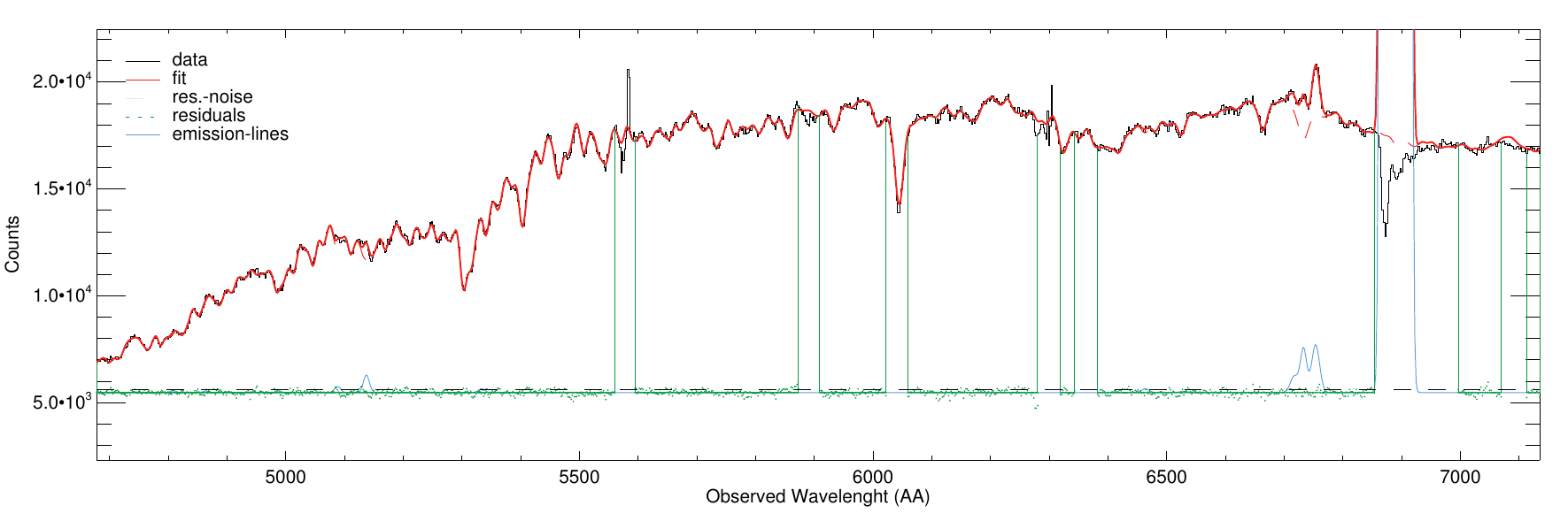}  
\caption{Reduced data and best-fit model of the central spectrum of NGC\,5228 using pPXF and GANDALF. The observed heliocentric velocity and dispersion of this galaxy is 7500 and 215 \kms, respectively. Strong sky lines (i.e. 5577 and 6300 \AA) and telluric features (6900 \AA) are masked from the fit. Emission from [OIII], [NII] and H$_\beta$ is detected in this spectrum. \todo{annotate the spectral features in figure.}} 
\label{fig:spectrum}
\end{figure*}}

\newcommand{\plotsdssbat}{\begin{figure*}
\centering
\includegraphics[width=0.6\textwidth,trim=0 0 0 0.0cm,clip=true]{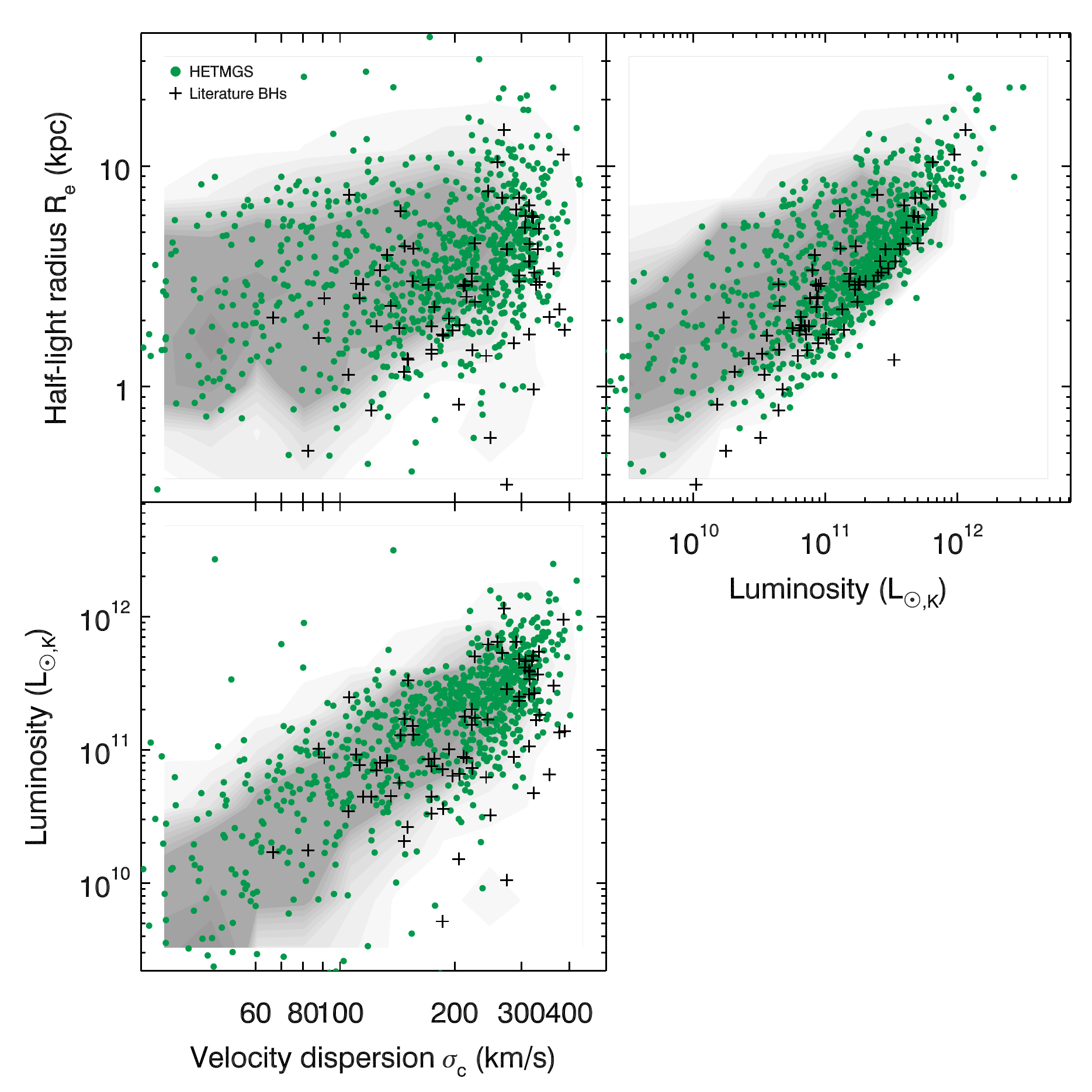}  

\caption{Demographics of the HETMGS galaxies and the literature black hole host galaxies shown as green circles and black crosses, resp. The shaded background represents number density of the global galaxy population, based on the representative SDSS sample of nearby galaxies. The \textbf{top left} panel shows the velocity dispersion versus half-light radius. The \textbf{right} panel shows the total luminosity versus half-light radius. The \textbf{bottom} panel shows the velocity dispersion versus luminosity, i.e. the Faber-Jackson relation. The HETMGS covers all of the galaxy parameter space, whereas the galaxies with a known black hole mass are not representative.} 
\label{fig:Sample}
\end{figure*}}

\newcommand{\plotcentralsig}{\begin{figure}
\centering 
\includegraphics[trim=0 0 11.8cm 0.0cm,clip=true,width=0.85\columnwidth,height=0.85\columnwidth]{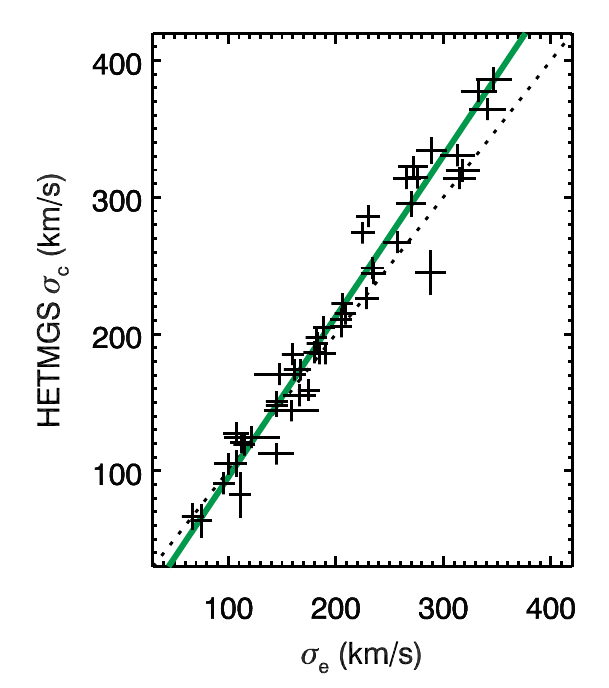} 

\caption{Comparison between the central dispersion $\sigma_c$ and the luminosity weighted dispersion inside the
half-light radius $\sigma_e$ for the subset of \HETNknownbh\ galaxies with black hole masses that are in the
survey. A linear regression shows that \centralsigvssige\ \kms. The slope is not consistent with unity. For high dispersion galaxies ($\sigma_c > 125$ \kms),
the central dispersions are higher than the half-light dispersions. See~\S\ref{sec:sigmaevssigmac}. 
}
\label{fig:centralsig} \end{figure}}

\newcommand{\plothettime}{\begin{figure*}
\centering
\includegraphics[width=0.90\textwidth]{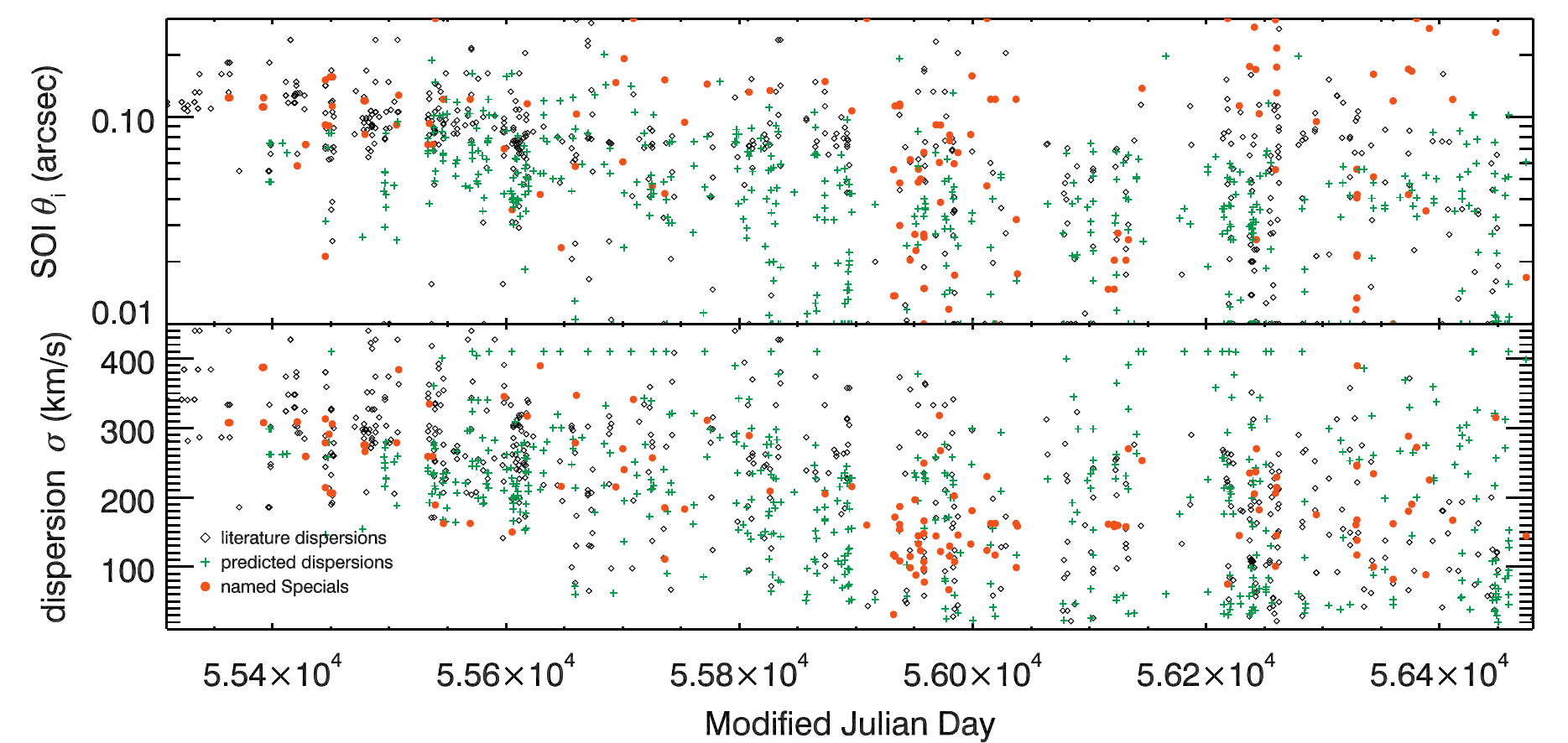}  
\caption{The \textbf{top} and \textbf{bottom }panels show the \SOI\ and velocity
  dispersion of the observed targets as a function of Modified Julian
  Day. The diamonds denote the observed targets with a literature
  dispersion from the final parent sample. When no literature
  dispersion is available a predicted velocity dispersion is indicated
  by a cross (Capped at 410
  \kms. See~\S\ref{sec:dispersion_estimates_from_fundamental_plane}.).
  Special galaxies (see \S~\ref{sec:special_targets}) are shown as
  filled circles. As the survey progressed, galaxies were targeted
  with smaller dispersions and \SOIs .}
\label{fig:hettime}
\end{figure*}}

\section{HET Massive Galaxy Survey}
 
The masses of black holes, \Mbh, correlate to various properties of their host galaxies. These correlations are the foundation for theories of the (co-)evolution of supermassive black holes and their host galaxies.  See \cite{2013ARA&A..51..511K} for a recent review. \referee{\protect The most commonly used black hole scaling relations are with stellar velocity dispersion \citep[\Msigma;][]{2000ApJ...539L..13G,2000ApJ...539L...9F}, bulge mass \citep[\Mbulge;][]{2004ApJ...604L..89H}, bulge luminosity \citep[\MLrel;][]{2011MNRAS.413.1479S}, and total luminosity \citep[\MLtot;][]{2014ApJ...780...70L}.}\ Theories on the existence of these scaling relations range from causal links through direct feedback between the black hole and its host \citep{1999MNRAS.308L..39F} to a non-causal origin from random hierarchical galaxy-galaxy merging \citep{2011ApJ...734...92J}. Another possibility is that the black hole scaling relation only holds for elliptical galaxies and bulges \citep{2011Natur.469..374K}. However not enough black hole masses have been measured to place firm constraints on these hypotheses. Also note that the \ML\ and \Msigma\ not mutually consistent for the biggest black holes \citep{2007ApJ...670..249L}.   The main issue, besides small numbers, is that the host galaxy properties only span a small physical range, which is too small to discern between different scenarios, due to measurement uncertainties and the large intrinsic scatter in these relations. 

In addition, the black hole masses of nearby galaxies are critically important for black hole mass determinations at higher redshift. The mass measurements from quasars and active galactic nuclei (AGNs) are measured using reverberation mapping (and its secondary methods), which rely on an empirical calibration that assumes that active black holes follow the local potentially biased \Msigma\ relation \citep{2010ApJ...716..269W}.

Very few galaxies are close enough for direct black hole mass measurements. For a successful measurement, the central region of the galaxy where the potential of the black hole dominates needs to be spatially resolved ($\RoI$ $\equiv G$\Mbh$\sigma^{-2}$, see \S~\ref{sec:the_soi_criterium}). Even for the closest galaxies, the \SOI\ is typically much smaller than 1\arcsec. Hence, black hole masses can only be measured by telescope with the highest spatial resolution, such as the \emph{Hubble Space Telescope} (\emph{HST}), large ground-based telescopes assisted by adaptive optics, and radio facilities such as the Very Long Baseline Interferometry Network.

In order to make best use of high spatial resolution facilities, the potential targets for future black hole mass measurements need to be carefully chosen. However, most nearby galaxies (with distances $<140$ Mpc) do not have any spectra available, without which it is impossible to ascertain their suitability for expensive high spatial resolution follow-up. Even if spectra are available, the amalgamated literature in the HyperLeda database \citep{2003A&A...412...45P} is often unreliable (see \S~\ref{sec:kinematics_comparison}).

The HET Massive Galaxy Survey (HETMGS) aims to find those galaxies that are suitable for a dynamical black hole mass measurement. The survey was a large program on the Hobby-Eberly Telescope \citep[HET;][]{1998SPIE.3352...34R} that operated for 10 trimesters and obtained spectra of \HETNGals\ galaxies. While the spatial resolution of the HET observations are typically not good enough to measure the black hole masses, this survey lays the necessary groundwork for systematic black hole mass measurement campaigns. As a by-product, the survey also provides a census of the nearest galaxies.

This survey is ideally suited for the unique design of the Hobby-Eberly Telescope. The 10-meter mirror is spherical and has a fixed elevation angle of 55\dgr. The primary mirror is stationary during observations and objects are tracked by moving the optics and instruments in prime focus. The telescope can only point to a fraction of the sky at any given time. The visible `doughnut' is 8.4\dgr\ thick. As a target passes through the donut, the telescope can track it for approximately one hour, depending on the declination of the object. The telescope design makes it most suited for programs that have short exposures and have many targets distributed across the sky. All observations are executed by night operators to optimize the science output. The scheduling algorithm is highly flexible and continually ranks all objects in the queue by score from the time allocation committee, visibility, observing conditions, etc. This allows the resident astronomer to adapt to current conditions in real-time. For details see \cite{2007PASP..119..556S}.

This paper starts with the description of the survey, sample construction, and completeness metrics in \S~\ref{sec:selection_method}--\ref{sec:special_targets}. Then, the data reduction is described in \S~\ref{sec:marcario_low_resolution_spectrograph_data_reduction}. The primary data product is the stellar velocity dispersion since this is required to predict the \SOI. Subsequently, these dispersion measurements are validated in \S~\ref{sec:kinematics_comparison}. When no literature dispersion is available a priori, the \SOI\ can be estimated from the photometry using the fundamental plane that relates galaxy size, surface brightness, and velocity dispersion, and a custom fit of this fundamental plane is presented in \S~\ref{sec:dispersion_estimates_from_fundamental_plane}. The global properties of the HETMGS sample are described in \S~\ref{sec:HETgalaxies}. Then, in \S~\ref{sec:sigmaevssigmac}, we explore whether the galaxy's central velocity dispersion or an average velocity dispersion within the galaxy's half-light radius correlates more strongly with black hole mass. In \S~\ref{sec:black_hole_scaling_relations} we compare the black-hole--host-galaxy demographics with the HETMGS sample and show that the host galaxies currently on the scaling relations are strongly clustered and biased toward the densest galaxies. We also look at the distribution of galaxies with large \SOIs , and find that there are very few galaxies for which an under-massive black hole could be detected. Finally, we conclude in \S~\ref{sec:conclusions} with a discussion of future prospects.

\section{Survey Design}  
\label{sec:selection_method}
\plothettime

The survey ran from April 2010 to August 2013 (when the telescope was taken down for a major upgrade for HETDEX). The total survey consists of \HETNGals\ galaxies, with 1265 long-slit spectra taken over 550 hours. Initially, we focused on galaxies with the largest dispersion. As the survey progressed the criteria on target selection were gradually relaxed and targets with lower dispersions and \SOIs\ were added to the queue (\S~\ref{sec:msigmasel}). This time dependence is shown in Figure~\ref{fig:hettime}. The target selection was continually expanded and improved over the survey life time. Below, we summarize the general overview and strategy, and in the subsections that follow, we describe the target selection in detail. \S~\ref{sec:Parentsample} details the parent sample used to select targets. In particular, \S~\ref{sec:the_soi_criterium} describes the adopted \SOI\ metric, which defines the suitability for a black hole mass measurement. The main metric of success for the survey is to find the galaxies with the largest predicted \SOIs , and we quantify this in \S~\ref{sec:msigmasel}. Subsequently, we detail the success of the secondary goal of sampling all possible host galaxy properties in \S~\ref{sec:sampling_all_sizes}. Lastly, the survey also includes special galaxies, such as those with previously measured black hole masses. These galaxies are described in \S~\ref{sec:special_targets}. All observations and the derived galaxy parameters from the survey are presented in Table~\ref{tab:hetmgs}, and are also available from the project website at \url{\websiteurl}.

At the start of each new trimester, we submitted a fresh set of targets. Afterwards the target list was updated about once every 6 weeks in order to account for holes in Local Sidereal Time (LST) in our \emph{and} the overall telescope queue. New observations were processed as the survey progressed, which were then used to optimally select and schedule future targets. During the survey we gradually optimized the target selection and queuing. In the end, the algorithm was very sophisticated: it backfilled the queue with low priority objects in undersubscribed LSTs, while making sure that these new targets did not conflict with our own high priority targets. The algorithm also ensured that both rising and setting targets were available at any time, to work with pointing restrictions (e.g., limitations due to high wind). By always having targets available, our program could capitalize on undersubscribed times in the queue.

We do not select galaxies to observe based on their morphology. The only exception is to exclude strongly interacting galaxies and those that are too large to fit in the 8\arcm\ slit. All HETMGS galaxies are within the telescope's declination limit of $-11<\delta<73$\dgr, and we minimized target overlap with the SAURON \citep{2002MNRAS.329..513D}, \ATLAS\ \citep{2011MNRAS.413..813C} and \CALIFA\ \citep{2012A&A...538A...8S} surveys, as their integral field unit (IFU) observations are of a higher quality and publicly available. The observed sample is not complete since it was scheduled as a low priority filler program.

\subsection{Parent sample}
\label{sec:Parentsample}

The targets in the survey were selected from a (continually updated) parent sample which was constructed from 2MASS photometry and literature velocity dispersions. We chose the 2MASS Extended Source Catalog \citep[XSC;][]{2000AJ....119.2498J} as it is the best all-sky catalog of extended sources that is currently available. It is deep enough to include all the nearby galaxies for which direct black hole masses can be robustly measured. Additionally, the catalog contains many useful parameters such as half-light radius ($R_e$), position angle (PA), and flattening.

Note that the 2MASS observations are relatively shallow and the XSC is known to underestimate the total flux of apparently large, faint or low-surface-brightness objects by an appreciable amount \citep{2003AJ....125..525J,2013ApJ...764..151G, 2013ARA&A..51..511K, 2014ApJ...780...69L}. Furthermore, the half-light radii are measured over circular apertures and are not corrected for the effect of beam smearing. As a result many objects with a full width at half maximum (FWHM) smaller than 2\farcs5 are overestimated in size.\label{sec:2masscaveat}

The parent sample needs to include all possible targets, and must therefore be much larger than the survey itself. From the XSC all \HETtmassN\ objects were selected that have apparent $K_s$ magnitudes (\textsc{K\_M\_EXT}) within the range $5<m_{Ks}<11$. Fainter galaxies are expected to have a \SOI\ that cannot be resolved. For an apparent brightness of $m_{Ks}=11$ the largest expected \SoI\ in the parent sample is 0\farcs04 when adopting the \Msigma\ relation and the fundamental plane from \S~\ref{sec:dispersion_estimates_from_fundamental_plane}. Current state-of-the-art adaptive optics reaches 0\farcs1 FWHM and the $m_{Ks}=11$ limit of the parent sample is thus sufficient.

The parent sample is further augmented with velocity dispersions from various literature sources. We added the 7800 literature velocity dispersions that were available from the HyperLeda database in early 2011. The two biggest sources in that database are \cite{1995ApJS..100..105M} and \cite{1999MNRAS.305..259W}. Out of the 7800 galaxies, 4600 were not in our parent sample. (Most of these are fainter than $m_{Ks}>11$.) This increased the total size of the parent sample to \HETParentSample\ objects. We cross-matched the resulting catalog with SDSS NYU Value Added Catalog \citep{2005AJ....129.2562B, 2008ApJS..175..297A} which added another 4500 stellar velocity dispersions to the parent sample.  

We added redshift distances from the NASA Extragalactic Database (NED) and the 2MASS redshift survey \citep{2012ApJS..199...26H} and redshift-independent distances from NED-1D. For the redshift distances we use $H_0=70.5$ \kms\ from \cite{2009ApJS..180..330K} and the reference frame from \cite{2000ApJ...529..786M}, which corrects for Virgo infall, the Great Attractor, and the Shapley supercluster. Almost all galaxies in the parent sample have estimated distances. Out of the parent sample 12\% of the objects have no associated redshift, but this reduces to 0.7\% when objects near the galactic plane ($|b|<5$\dgr), large extinction ($A_b>2$) and non-galaxy-like colors ($0.75>J-K>1.25$) are removed \citep[see also][]{2012ApJS..199...26H}. For each object, the amount of Galactic foreground extinction was determined using the 100 $\mu\mathrm{m}$ dust maps from \cite{1998ApJ...500..525S}.

\subsection{The Sphere of Influence}  
\label{sec:the_soi_criterium}

For survey target selection, we need to adopt a criteria for determining the suitability of a galaxy for a black hole mass measurement. The mass of the black hole at the center of galaxies can best be determined by spatially resolving the kinematics near the black hole, using tracers like rotating gas or stars. The closer the tracer is to the black hole, the more accurate the measurement. Ideally, the region where the gravitational forces of the black hole dominates over the rest of the galaxy should be probed. The size of this sphere of influence is usually\footnote{An alternative to this dynamical definition is to use photometry. By deprojecting the photometry, the total amount of enclosed stellar mass as function of distance from the black hole can be estimated. The light can be converted into mass with a mass-to-light ratio (from either dynamics or a stellar population analysis), and the mass can then be compared to the mass of a putative black hole. However, the photometry needs to have sufficient spatial resolution to resolve the region of interest. Typical ground-based imaging is not good enough for this, and thus, for this survey, we adopt the the dynamical \SoI\ estimator.} defined as $\RoI$ $\equiv G$\Mbh$\sigma^{-2}$. Given distance $D$, the apparent size of the sphere of influence is then \SoI$=\RoI D^{-1}$. This definition is derived from the virial theorem by assuming a spherical geometry and isotropy near the black hole \citep{1972ApJ...178..371P}. Sometimes a less conservative definition is adopted that uses the diameter instead of the radius \citep[e.g.,][]{2004ApJ...602...66V}.

 The definition of \SoI\ is very convenient, especially when planning observations where the black hole mass is not yet known a priori. By adopting a black hole scaling relation, the \SOI\ can be estimated and the feasibility of proposed observations can be evaluated. The most practical scaling relation for this purpose is \Msigma, which yields a prediction for the \SOI\ \SoI$= G\sigma^{2.5}D^{-1}$, when adopting a slope of 4.5 in \Msigma\ from \cite{2009ApJ...698..198G}. As a result, the sphere of influence can be predicted with just a distance and a velocity dispersion as input. Note that \SoI\ is just a prediction; it depends on the adopted scaling relations and the true \SOI\ can only be known once the black hole mass has been measured. Throughout this paper \SoI\ is repeatedly used, and it always refers to the \emph{predicted} \SoI, unless the black hole mass has been measured.

The \Msigma\ relation is the only practical black hole mass proxy available. Other scaling relations have larger intrinsic scatter or require a more detailed analysis of the host galaxy. For example, the \MLrel\ relation relies on bulge--disc decompositions. As such the resulting measurement of bulge luminosity is often very degenerate \citep{2014ApJ...780...69L} and cannot be robustly determined with an simple algorithm that can be applied to a large data set. Hence it is impractical to use the bulge luminosity to select targets. On the other hand, the velocity dispersion $\sigma$ can be measured cleanly. Furthermore, the \Msigma\ relation also yields a smaller---more conservative---estimate of \SoI\ than expected from the \MLtot\ relation (See \S~\ref{sec:soi_distribution}). Many other scaling relations exist, \citep[e.g.,][]{2007ApJ...665..120A, 2012ApJ...746..113G, 2007ApJ...669...67H, 2012ApJ...746..113G}, but those are not a significant improvement of \Msigma. In practice, such alternative predictors would have targeted much the same galaxies. See \S~\ref{sec:black_hole_scaling_relations} for a direct comparison with \MLtot.

\subsection{Sphere of influence completeness}  
\label{sec:msigmasel}
\plothetsois

The survey originally started by targeting the 65 galaxies with highest literature dispersions $\sigma > 270 $ \kms\ and \SOIs\ $\theta_i>$0\farcs1 solely based on the HyperLeda database. We adopt the \Msigma\ parameters from \cite{2009ApJ...698..198G} to estimate the \SOI. As the survey progressed and transitioned into a large program, these hard limits on $\sigma$ and \SoI\ were gradually relaxed and additional sources were added (Figure~\ref{fig:hettime}).

The \SOI\ of the \HETNwithdisp\ targeted galaxies \emph{with} literature dispersions are shown in left panel of Figure~\ref{fig:hetsois}. Many of these literature dispersions turn out to be overestimated compared to the HETMGS measurements. Of the \Ngalbigsoisig\ observed galaxies with a predicted $\theta_i>$0\farcs1 a significant percentage (\Ngalbigsoisigfrac\%) turned out to have $\theta_i<$0\farcs08. This is largely caused by the large uncertainties and systematics of the HyperLeda velocity dispersions, which is also apparent in Figure~\ref{fig:hetvsvagcvshl}.

The middle panel of Figure~\ref{fig:hetsois} shows the \SOI\ of all the \HETNwithoutdisp\ observed targets without literature dispersions. For the galaxies that did not have a literature velocity dispersion, the \SOI\ was predicated by combining the \Msigma\ relation and a dispersion estimate from the fundamental plane from \S~\ref{sec:dispersion_estimates_from_fundamental_plane}. These dispersion estimates are good enough to yield adequate predictions for $\theta_i$. When selecting targets this way, we omitted probable quasars that have both small sizes ($r_e<2$\arcsec) and AGN-like colors ($0.75<J-K<1.25$). In total, we found \HETNwithbigsoifromfp\ objects with \SOIs\ bigger than 0\farcs1 that also did not have a velocity dispersion measurement in the literature.

The final survey include 95\% and 50\% of all galaxies with a \SOI\ above \SOIcomplete \arcs\ and \SOIcompletefif \arcs, respectively (from the subset of galaxies in the parent sample, visible from the HET, and excluding galaxies with a known black hole mass and \ATLAS\ data). The right panel of Figure~\ref{fig:hetsois} shows the predicted \SOIs\ of all the HETMGS observed galaxies based on our dispersions. The final sample contains \HETNwithbigsoi\ galaxies with a \SOI\ greater than 0\farcs1. Under the assumption that \Msigma\ is a reasonable predictor for black hole mass, the survey has satisfied its goal of finding almost all galaxies with a large \SOI in the parent sample.

The \SoI\ is strongly coupled to galaxy dispersion ($\theta_i \propto \sigma^{2.5}$) and galaxies with large dispersions have the largest \SOI. Selecting galaxies based on largest \SoI, as is done here, is strongly biased towards galaxies with the largest densities. This is apparent as the strong clustering in Figure~\ref{fig:2masspf}. Many object in the survey are denser still than average early-type galaxies.

\subsection{Sampling all sizes and surface brightnesses}  
\label{sec:sampling_all_sizes}
\plotHETfp  

If targets were selected merely on the largest possible \SoI\ using the \Msigma\ relation, the survey would contain mostly dense (early-type) galaxies with dispersions over 200~\kms, as these galaxies have the largest \SOIs . This bias occurs because the most important term in \SoI\ is $\sigma^{2.5}$. This is directly visible in figure~\ref{fig:hetsois}, as the highest dispersion galaxies have the largest \SoI. The early-type galaxies are already over-represented in black hole studies and hence diversifying the parameter space to other types of galaxies would create leverage on the scaling relations. Especially to test the universality  (e.g., spiral galaxies appear to have low black hole masses for their dispersions; \citealt{2010ApJ...721...26G, 2011ApJ...727...20K}). 

To ensure that the survey is in not solely restricted to high dispersion galaxies, we need to employ an alternative way of selecting targets, independent from velocity dispersion. A good alternative would be to use use bulge luminosities, as that correlates very well with black hole mass also. However, the bulge luminosity is impractical; they require a photometric decomposition, to separate the disk and the bulge. Those decompositions are often very degenerate and require a lot of fine-tuning. Other desirable galaxy parameters, like optical ($g$-$r$) colours, are not available over the whole sky and can thus not be used when the distance (in Mpc) of the targets is key priority.  

We chose to select additional targets by sampling in galaxy size and average surface brightness (\Amu). These two properties make up a near face-on projection of the fundamental plane ($R_e \propto \sigma_e^\alpha \Amu^\beta$, see~\S\ref{sec:dispersion_estimates_from_fundamental_plane}) and is also independent of the (often unknown) velocity dispersion. For our purposes, surface brightness is a better choice than absolute luminosity, as the latter has a much tighter relation with galaxy size. This selection automatically ensures sampling across galaxy types, because Spiral galaxies occupy a different region on fundamental plane than early-type galaxies.

For a given set of galaxies with the same properties, the ones with the largest \SoI\ will have the lowest distances (in Mpc). To diversify the sample in size $R_e$ and surface brightness $\Amu$, we only have to do a sampling in these two parameters and select the closest ones (in Mpc) at every given value-pair of $R_e$ and $\Amu$. We define a euclidian distance norm of $||x||:=\sqrt{\Delta \Amu^2+\Delta \log(R_e)^2}$, with \Amu\ in units of mag$/$arcsec$^2$ and $R_e$ in kpc. This allows us to quantify our sampling in the $R_e$--$\Amu$ plane. In theory the strategy would be straightforward: Just select the nearest galaxies, until sufficient coverage is created. In practice, the targeting is dependent of what galaxies can be observed at the telescope in the remainder of the trimester. Hence we created an algorithm that would add the nearest (in Mpc) galaxies to our sample to would best enhance our coverage w.r.t the norm. This used an iterative method; Every time new targets were scheduled, we sorted the subsample of visible targets in order of increasing distance in Mpc. Then, we started with the nearest galaxy and checked if it had any nearer (in Mpc) neighbors with norm $||x|| < \epsilon$ among the previously observed and queued objects still visible in the remainder of the trimester. If not, then the galaxy is added to queue. As more galaxies were observed, we decreased $\epsilon$ in order to find enough observable targets. At the end of the survey $\epsilon$ was reduced to 0.07. The method was also the primary method for backfilling the undersubscribed LST slots in the telescope queue, by iterative shrinking $\epsilon$ until targets where found with the required LST.

This method is completely independent of the adopted black hole mass scaling relation, and it produces a survey sample that has the same extent in galaxy size and surface brightness as the parent sample (see Figure~\ref{fig:2masspf}). Furthermore, the selection method picks out the closest galaxies with the largest \SOI\ for a set of galaxy properties. In the final survey, 90\% of the galaxies have a neighbor within a norm of $||x||<$\fpdissel. Our procedure selected many nearby spirals galaxies that otherwise would not be included in the survey at all, including the most extreme galaxies such as the biggest and smallest spiral galaxies. Many of the galaxies selected this way are the low \SOI\ objects seen in Figure~\ref{fig:hetsois}.

\subsection{Special targets}  
\label{sec:special_targets}

The survey also contains $\sim$150 special galaxies, which were targeted to create overlap with other black hole studies and allow for cross-calibration checks. These galaxies were drawn from a dedicated list, and were queued at an elevated priority. The red dots in the timeline in Figure~\ref{fig:hettime} represent the special galaxies, as reconstructed from the final list. Once observed, these objects were not treated differently in terms of data reduction or kinematic analysis, and are considered part of the total sample. All special galaxies are included in Table~\ref{tab:hetmgs} and the figures.

More specifically, the special targets include galaxies with direct black hole mass measurements from literature compilations \citep[and references therein]{2013ApJ...764..184M,2013ApJ...764..151G,2013ARA&A..51..511K} and black hole estimates from \cite{2009ApJ...692..856B}. We observed \HETNknownbh\ of the \Nknownbh\ Northern galaxies, with a known black hole mass. We also targeted galaxies for which black hole mass efforts are currently under way, but are yet unpublished (Greene, priv.\ comm., Krajnovi\'c, priv.\ comm., McConnell, priv.\ comm.). Most of those galaxies were already in the sample, and were previously selected, following the procedure outlined in \S~\ref{sec:the_soi_criterium}. The primary reason to observe these objects is to create a homogeneous data set among the galaxies with known black hole masses. In \S~\ref{sec:sigmaevssigmac}, we use this homogeneity to show that the central dispersion, $\sigma_c$, can be used in \Msigma.

One of the first results from the HETMGS is the discovery of compact galaxies that may host black holes weighing $\sim$10\% of their host galaxy mass \citep{2012Natur.491..729V, 2013MNRAS.434L..31L, WalshN1271, Yildirim1}. These \"uber-massive black hole candidates appear to be significantly more massive than expected based on their host galaxy's luminosity. The host galaxies are relatively small and have high velocity dispersions for their (stellar) mass. In our survey, such objects are selected because they have large \SOIs\ due to their high dispersions and because of the diversity sampling from \S~\ref{sec:sampling_all_sizes} that targets extreme galaxies. Nevertheless, we specifically targeted compact galaxies using a mass--size cut: $-6.91 - 0.29 M_{Ks} > \log(0.75 R_e)$. We further required that the galaxy have no previous velocity dispersion measurement, or a literature velocity dispersion value that was 20\% higher than predicted from our fundamental plane calibration (see \S~\ref{sec:dispersion_estimates_from_fundamental_plane}). This selection criteria produced about 30 targets. Overall, there are 200 galaxies below this mass--size cut in the survey. A full analysis and discussion of these compact galaxy systems is outside of the scope of this paper.

\section{Marcario Low Resolution Spectrograph data reduction}  
\label{sec:marcario_low_resolution_spectrograph_data_reduction}
\plotspectrum 

All the observations were taken with the Marcario Low Resolution Spectrograph \citep[LRS;][]{1998SPIE.3355..375H}, which is an optical long-slit spectrograph with a slit length of 4\arcm\ and a 3k$\times$1k CCD. We used the g2 grating, which covers 4200--7400 \AA, and the default 2$\times$2 binning. This setup provides an instrumental resolution of 4.8 \AA\ (7.5 \AA) FWHM, or a dispersion of 108 \kms (180 \kms) for the 1\arcsec--wide (2\arcsec--wide) slit, as measured from the 5577 \AA\ night sky line. When practical, we aligned the slit on the major axis and centered it on the galaxy. During each visit, we obtained a single 15 minute exposure. The typical spatial resolution of the observations is 2\farcs5 FWHM.

The observations and calibration data were retrieved and processed using a dedicated pipeline written in IDL\footnote{We gratefully acknowledge the use of the IDL astrolib \citep{1993ASPC...52..246L} and MPFit \citep{2009ASPC..411..251M}}. The pipeline is fully automated and executes a series of basic steps. First, the spurious pixels in each exposure are masked using a bad pixel map, and cosmic rays are masked using PyCosmic \citep{2012A&A...545A.137H}. The frames are overscan and bias subtracted, and then flat fielded. The next step is a direct interpolation onto a frame that is linear in the spatial dimension and logarithmic in the wavelength dimension, using a spatial distortion map from a pinhole exposure and the wavelength solution from nightly arc lamp exposures. A variance (error) frame is propagated in the same way. During each exposure, the effective aperture of the HET continually changes because the prime-focus optics move to track an object, thereby affecting the throughput along the spatial direction. These throughput changes are measured in each science exposure using the apparent brightness of the sky lines along the slit. For this reason, we do not attempt to perform absolute or relative flux calibration\footnote{The LRS calibration program does provide flux and radial velocities standards stars on a regular cadence.}, and no attempt was made to measure the absolute emission-line fluxes.

The data products, presented in Table~\ref{tab:hetmgs}, include the following quantities: heliocentric velocity, stellar absorption line velocity dispersions, the emission line ratios [\ion{N}{2}]/H$\alpha$ and [\ion{O}{3}]/H$\beta$, and their uncertainties. These quantities are all measured within a 3\farcs5 aperture centered on the brightest pixel in the slit. For the stellar kinematics, the pipeline fits the stellar continuum with the pixel-fitting code \citep[pPXF;][]{2004PASP..116..138C} using template stars from MILES \citep{2006MNRAS.371..703S,2011A&A...532A..95F}. Inside the central aperture the Signal-to-Noise ratio (S/N) is typically over 100, resulting in robust stellar kinematic measurements, as shown in \S~\ref{sec:kinematics_comparison}. The fit to each (central) galaxy spectrum is manually inspected to find systematics and errors. An example spectrum is shown in figure~\ref{fig:spectrum}.

The stellar kinematic extraction is obtained from the stellar continuum in an observed window of 5000 -- 6100 \AA, selected to minimise instrumental resolution changes across the slit. Additionally all regions with possible contamination from emission lines were masked. For the both the stellar kinematics and emission-line extraction, we apply multiplicative (\textsc{mdegree}) and additive polynomials (\textsc{degree}) of degree 25 and 5, respectively, to the template stars to match the continuum and account for dust, AGN light and the (absence of) flux calibration. \referee{The number of polynomials is chosen to be large enough to be able to compensate for (uncommon) issues with the uneven flat fields. In the next section, we show that our kinematics are robust.}

The sky is measured in two large bins on each end of the slit, and were added as a sky-template to pPXF to do the sky subtraction. For the emission-line fit we use all the 1000 template stars in the MILES library and the full wavelength range available. For the stellar kinematics, we construct a single composite template by first fitting the full wavelength range using a subset of 90 stars that get used most frequently in all the aforementioned emission line extractions. Adding more stars increases the computational time and does not change the derived values. Typically 20 stars get a non-zero weight in the initial pPXF fit. The uncertainties on the stellar kinematics were computed by running a 100-iteration Monte-Carlo in which the flux is randomly perturbed and with a single composite template. During the Monte-Carlo the instrumental dispersion of the composite template is also perturbed, which insures realistic uncertainties when the galaxy dispersion is unresolved.  The mean and standard deviation of the Monte-Carlo run are adopted as the measured value and the formal 1$\sigma$ uncertainty.

When present, emission lines are obtained using GANDALF \citep{2006MNRAS.366.1151S}. An example GANDALF fit is shown in figure~\ref{fig:spectrum}. We consider an emission as detected when the residuals underneath the line are 4 times lower than the amplitude of the line \citep[Amplitude-over-Noise AoN$>$4; see][]{2006MNRAS.366.1151S}. The errors on the emission lines are directly computed by GANDALF. The [\ion{S}{2}] emission lines often overlap with a broad telluric feature, and therefor are not included in Table~\ref{tab:hetmgs}.

The survey contains \HETNagn\ galaxies with broad emission lines, which are presumably AGN. The dispersion for these objects could not be reliably measured in the central aperture, as the stellar continuum is washed out. Instead, an attempt was made to measure the dispersion using a larger aperture and masking the nucleus. Such objects are flagged in Table~\ref{tab:hetmgs} as AGN.

\begin{figure}
\centering
\includegraphics[trim=0 0 0 0.45cm,clip=true,width=0.95\columnwidth]{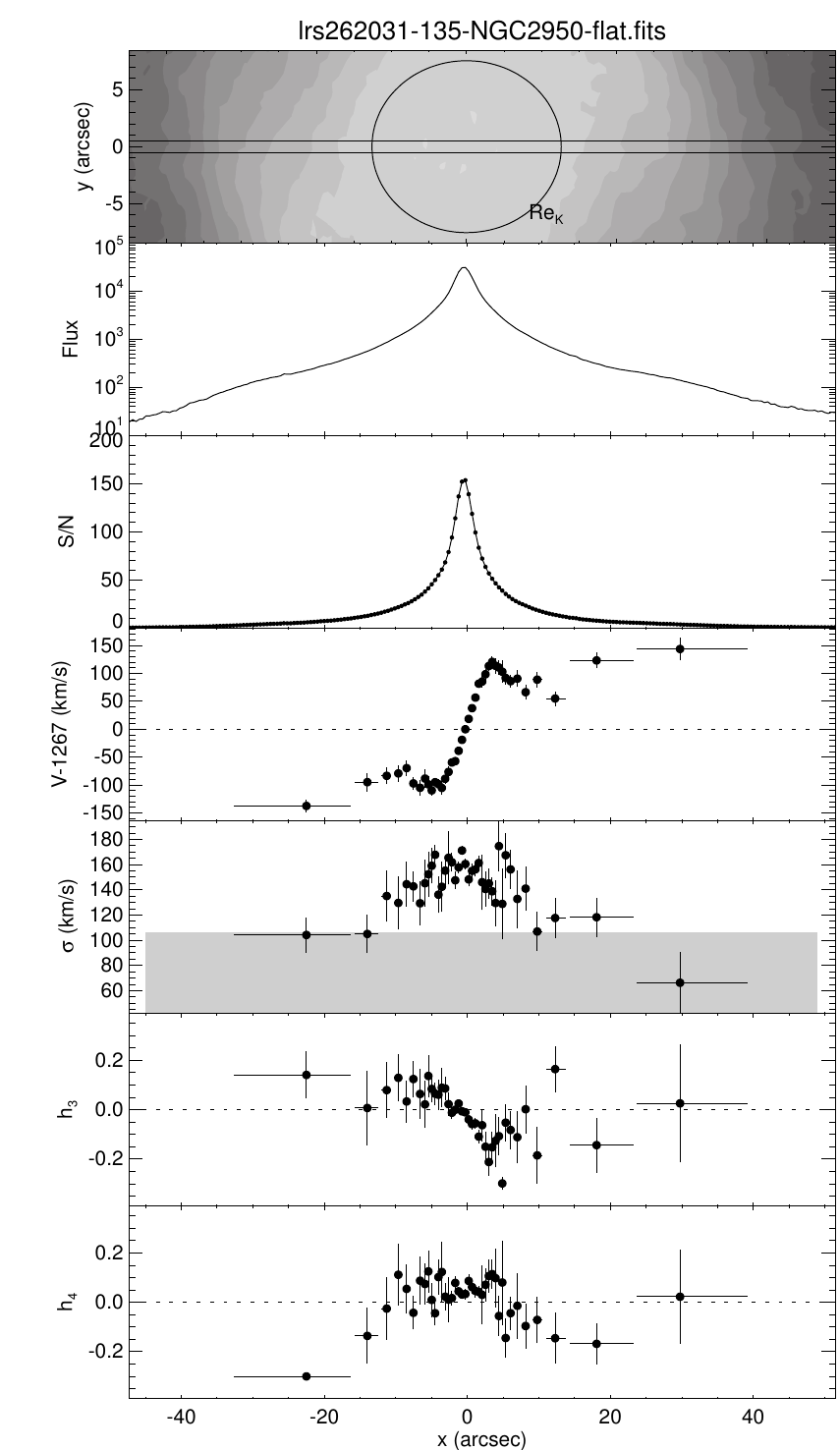}

\caption{Example of the resolved stellar kinematics using  exposure 262031 of NGC2950. The kinematics are measured in bins that are combined to reach Signal-to-Noise of 35. Top to bottom panels show: Finder chart and approximate slit alignment based on the DSS image. (Unitless) Flux. S/N. Stellar velocity, dispersion, and Gauss-Hermite moment h$_3$ and h$_4$. The grey shaded region indicates where the instrumental dispersion is larger than the stellar velocity dispersion. }
\label{fig:kinematics}
\end{figure}

Most observations were deep enough that spatially resolved kinematics could be extracted as well, allowing us to measure rotation curves and velocity dispersion profiles. To ensure adequate signal for the stellar kinematics extraction, we combine spatial rows into bins with a minimum S/N of 25. An example is shown in figure~\ref{fig:kinematics}. The kinematics are available from the project website at survey website \url{\websiteurl}.

\section{Kinematics Comparison}  
\label{sec:kinematics_comparison}
\plothetvscagcvshl 

There is a large overlap between the HETMGS and other spectroscopic surveys, which we use to validate our heliocentric velocities and stellar velocity dispersions. The SDSS survey provides a good comparison, as these homogeneous measurements all come from the same instrument and its 3\arcsec\ fibers are approximately the same size as our pseudo-apertures from the HET long slit. There are \HETSDSSoverlap\ galaxies in common with SDSS and there is excellent agreement with their dispersion measurements, as shown in Figure~\ref{fig:hetvsvagcvshl}. The standard deviation of the relative difference between the two dataset is only \HETSDSSstddev\%. Which is smaller than the \HETrelstddevOfSdsssubset\% relative uncertainty of the HETMGS errors of the same subset. Such small differences are consistent with the expected systematics due to the changes in seeing and positioning of the apertures of the different telescopes.

The velocity dispersions from HyperLeda are much less reliable, as this database consists of amalgamated literature. The \HETHLoverlap\ galaxies in common have a standard deviation in the difference of 16\% or $30$ \kms. It is noteworthy that HyperLeda values over 300 \kms\ are commonly too large by a significant amount, as seen in Figure~\ref{fig:hetvsvagcvshl}. There is an overlap of \SDSSHLoverlap\ galaxies between SDSS and HyperLeda, and the comparison between the two data sets shows the same trend. These biases and large uncertainties need to be taken into account when preparing observations with HyperLeda.

The Palomar survey by \cite{2009ApJS..183....1H} published velocity dispersions for the 428 galaxies with the largest apparent brightnesses. These values were not yet included in the HyperLeda database when the parent sample was constructed. There are 188 galaxies that overlap for which were find agreement that is similar to that of the HyperLeda sample: the standard deviation of the difference is 17\%.

The SDSS redshifts are very precise with a formal uncertainty of \SDSSvelformerror~\kms. In the cross-matched sample between HETMGS and SDSS the mean difference between the heliocentric velocities is \HETSDSSVELmean\ \kms\ with a standard deviation of \HETSDSSVELstddev\ \kms. The redshift comparison with the rest of the parent sample shows a larger offset: their mean and standard deviation are $\HETVELmean$ and \HETVELstddev\ \kms, respectively. In rotating galaxies the mean velocity changes rapidly near the center, and thus these kinematic offsets are expected just based on small positional offsets.

\section{The Fundamental plane as dispersion predictor}  
\label{sec:dispersion_estimates_from_fundamental_plane}
\plotsigmapred

For the target selection we need an estimate of \SoI. However most of the galaxies in the parent sample do not have a literature stellar velocity dispersion measurement, which is required to compute their \SoI. In those cases the fundamental plane $R_e \propto \sigma^\alpha \Amu^\beta$---the relation between galaxy size, surface brightness and dispersion---can be employed to estimate their dispersion. To get dispersion estimates, we fitted a fundamental plane to the combination of XSC photometry and the SDSS and HETMGS dispersions. After some experimentation, the best results were achieved with the 2MASS catalog values for the half light radius $R_e$ (\textsc{K\_R\_EFF}) and effective surface brightness ($\Amu=\mathrm{K\_M\_EXT}+A_K+2.5 \log(2\pi R_e) +0.394$) combined with the dispersion from the HETMGS and the SDSS galaxies within 140 Mpc. With this dataset, a robust quartic fit from \textsc{ROBLIB} yields a fundamental surface that predict the velocity dispersion of all the galaxies in our parent sample with an RMS error of 0.09 dex. The residuals, shown in figure~\ref{fig:sigmapred}, do not show systematic biases, which is remarkable as the sample contains both early and Late type galaxies. And the fundamental plane is normally only used for early types. Our fundamental plane parameters are a significant improvement over the parameters from \cite{1998AJ....116.1591P}, which has an RMS error of 0.2 dex on this data set, and only includes ETGs.

A flat plane fit resulted in larger residuals and a systematic bias for the largest dispersion galaxies, which is unfortunate as they are the highest priority targets of the HETMGS. The curvature in our quartic surface is not very strong, as seen in figure~\ref{fig:2masspf}. \referee{\protect The coefficients of the surface fit are \quarticsurfacecoef .}

%$\log\sigma = a+b \Amu +c \log R_e +e\Amu^2+e (\log R_e )^2+f\Amu \log(R_e)$

For the survey, only the predictive power of the fundamental plane is important. The functional form is irrelevant, as long as the dispersion is predicted reliably. Nonetheless, it is remarkable the fundamental plane derived from photometry of the 2MASS XSC catalog has so little scatter. A full treatment of a 2MASS fundamental plane is outside of the scope of this paper, but see \cite{2012MNRAS.427..245M} for a near-infrared study based on $10^4$ early types from 6dFGS, and also see \cite{2011MNRAS.417.1787F} for a near-IR fundamental plane that includes both early- and late type galaxies from the SAURON survey.

During the target selection process, the dispersion estimated from the fundamental plane is used when no literature dispersion measurement is available. Velocity dispersion predictions over 410~\kms\ were truncated, as such galaxies are not expected to exist, nor found in our survey. In practice most of these objects turn out to be quasars.

\section{The HETMGS galaxies}
\label{sec:HETgalaxies}
\plotsdssbat  

With \HETNGals\ galaxies, the HETMGS survey is the largest galaxy survey with spatially resolved spectroscopy to date. The final sample has a median distance of \HETmeddist~Mpc and all available combinations of galaxy size, luminosity, and dispersions, as shown in figure~\ref{fig:Sample}. The survey is representative of the local volume above a M$_k < -21$, and sizes of 0.5 kpc. The survey is strongly weighted towards the densest galaxies. This is apparent in the mass-size panel. Those objects have the highest dispersions, which is the most important factor for the \SOI\ ($\SoI \propto \sigma^{2.5}/D$). The sample contains about 30\% late type galaxies, based on the galaxy morphology identifiers from HyperLeda.

There is a significant amount of overlap between the HETMGS and other surveys. Between \ATLAS, \CALIFA\ mother-sample, SDSS and the Palomar survey, there is 56, 83, \HETSDSSoverlap\ and 188 galaxies overlap. However all these samples are very different. For example, the Palomar spectroscopic survey \citep{2009ApJS..183....1H} contains 426 velocity dispersions for the brightest galaxies with apparent magnitudes of $B < 12.5$. The SDSS only includes a (small) fraction of the nearby galaxies \citep{2008ApJS..175..297A}. And the 600 \CALIFA\ galaxies are selected within $0.005<z<0.03$ and are typically too far away for a black hole mass measurement \citep{2012A&A...538A...8S, 2014arXiv1407.2939W}. The 260 \ATLAS\ \citep{2011MNRAS.413..813C} galaxies includes all the massive early-type galaxies inside 24~Mpc, but misses large luminosity galaxies due to cosmic variance. The comparison of galaxy properties between HETMGS and \CALIFA\ and \ATLAS\ is shown in Figure~\ref{fig:2masspf}.

Emission lines are quite common in the centers of galaxies, often caused by star formation, shocks or accreting black holes. These phenomena can occur at the same time \citep[e.g.][]{2013A&A...558A..43S}. In one third of the survey, central emission lines are detected. This includes the \HETNagn\ broad line AGNs, and 203, 140 and 290 objects of that are classified as Seyfert, LINER and HII star formation, according to the classification scheme devised by \cite{2006MNRAS.372..961K}. But note that we do not specifically attempt to detect (weak) broad line components with the GANDALF. The AGNs can vary substantially on both long and short time scales. One such dramatic example, reported in \cite{2014arXiv1404.4879D}, is Mrk~590 in which the broad lines have all but disappeared. The HETMGS spectra could be used as baseline for other synpotic AGN studies. 

\section{Central dispersion or average dispersion?}
\label{sec:sigmaevssigmac}
\plotcentralsig

What correlates better with black hole mass: the central dispersion $\sigma_c$ or the luminosity weighted dispersion inside a half-light radius $\sigma_e$? The first one is more easily measured, but probes an seemingly arbitrary scale of the galaxy, that depends on distance and instrumental effects. Whereas $\sigma_e$ is measured\footnote{Note that method for measuring $\sigma_e$ varies; the measurement requires integral field observations to get the 2D luminosity weighted dispersion and that is not always available. Also the definition of half-light radius varies; e.g. some authors only use the bulge $R_e$.} inside a physically relevant aperture ($R_e$) and is also the quantity used for the fundamental plane and virial galaxy masses. Different apertures for $\sigma$ have been used in the literature. For example, \cite{2011Natur.480..215M} has suggested that the central region near the black hole should excluded from the $\sigma_e$ estimate, while \cite{2013ApJ...772...49W} argues for the use of a $\sigma$ that only contains half of the second moment. Furthermore the known black hole masses themselves are strongly correlated with distance, with bigger black holes being further away. Given that $\sigma_c$ is the most commonly available quantity it is useful to know if and how it correlates with black hole mass.

The difference between $\sigma_c$ and $\sigma_e$ is most interesting for disk galaxies where these two measurements probe very different things. Consider the broad-line AGNs in bulge-less disks galaxies \citep{2010ApJ...721...26G, 2013MNRAS.429.2199S, 2013ApJ...775..116R}. For these systems, \cite{2011Natur.469..374K} argues that neither \Msigma\ and \MLrel\ should apply to these galaxies, however their AGN is proof that they host a black hole. Is there perhaps another scaling relation that is applicable for these systems? If the black hole mass is solely linked to the bulge \citep[e.g.][]{2013ApJ...772...49W}, then one would expect that the $\sigma_e$ of a disk dominated galaxy should not correlate with black hole mass, whereas $\sigma_c$ should correlate better as it is a good tracer of the (central) bulge component. A homogenous dataset allows for a systematic test of all such scenarios.

For the purpose of the survey, we are predominantly interested in whether the $\sigma_c$ is a good predictor of black hole mass. \cite{2009ApJ...698..198G} found their dataset to be consistent with $\sigma_c=\sigma_e$, with an RMS scatter of 22 \kms. But this warrants repeating with the homogeneous dispersions from the HETMGS and the literature updates to $\sigma_e$ (e.g. \ATLAS). For this comparison we use $\sigma_c$ as measured in the central HETMGS 3\farcs 5 aperture (See \S\ref{sec:marcario_low_resolution_spectrograph_data_reduction}) and $\sigma_e$ as tabulated in \cite{2013ARA&A..51..511K} or \cite{2013ApJ...764..184M}. There are \HETNknownbh\ galaxies with a $\sigma_c$ and a black hole mass in this survey. The comparison is shown in figure~\ref{fig:centralsig}. Using a linear regression \citep[LINMIXERR,][]{2007ApJ...665.1489K}, we find \centralsigvssige\ \kms. The slope is inconsistent with being unity and is also inconsistent with the regression from \cite{2009ApJ...698..198G}. The $\sigma_c$ is higher for higher dispersion galaxies and lower for the disk-dominated galaxies with low dispersions.

Using the same subset of \HETNknownbh\ galaxies and LINMIXERR the \MsigmaE\ and \MsigmaC\ relations are: $(\alpha, \beta, \epsilon_0)=$ \msigmacoef, with $\log( M_\bullet/ $\Msun$ )=\alpha + \beta \log(\sigma/200 $\kms$)$, where $\epsilon_0$ is the intrinsic scatter. The two relations are nearly identical, apart from a slightly shallower slope with the central dispersion, as expected from the correlation between $\sigma_c$ and $\sigma_e$. So both velocity dispersion measures can be used to predict black hole mass. Our \MsigmaE\ relation is not consistent with the $\beta=4.38$ from \cite{2013ARA&A..51..511K}, because we did not exclude pseudo-bulges (see their \S6.6.2). Our \Mbh--$\sigma_e$ relation is consistent with \cite{2013ApJ...764..184M}.

We conclude that $\sigma_c$ can indeed be used as a substitute in \Msigma, albeit with a steeper gradient. \referee{\protect In the survey we used \Msigma\ from \cite{2009ApJ...698..198G} for the target selection (\S\ref{sec:msigmasel}), which has a shallower slope. The difference with  \MsigmaC\  is not very big, as seen in figure~\ref{fig:MLlimits}. If we had \MsigmaC\ for target selection instead, the survey would have favored higher dispersion galaxies at larger distances, as this relation predicts bigger black holes for the highest dispersion galaxies. However those galaxies were already included in the survey, as they already have the largest \SOIs\ anyways. The completeness statistics therefor would not differ appreciably with the \MsigmaC. The biggest change would be a decrease of  $\sim30$\% of the  \SoI\ of low dispersion galaxies in the survey ($\sigma_c < 150$~\kms). }

\section{Black Hole Demography}  
\label{sec:soi_distribution}

\begin{figure*}
\centering
\includegraphics[width=1.0\textwidth]{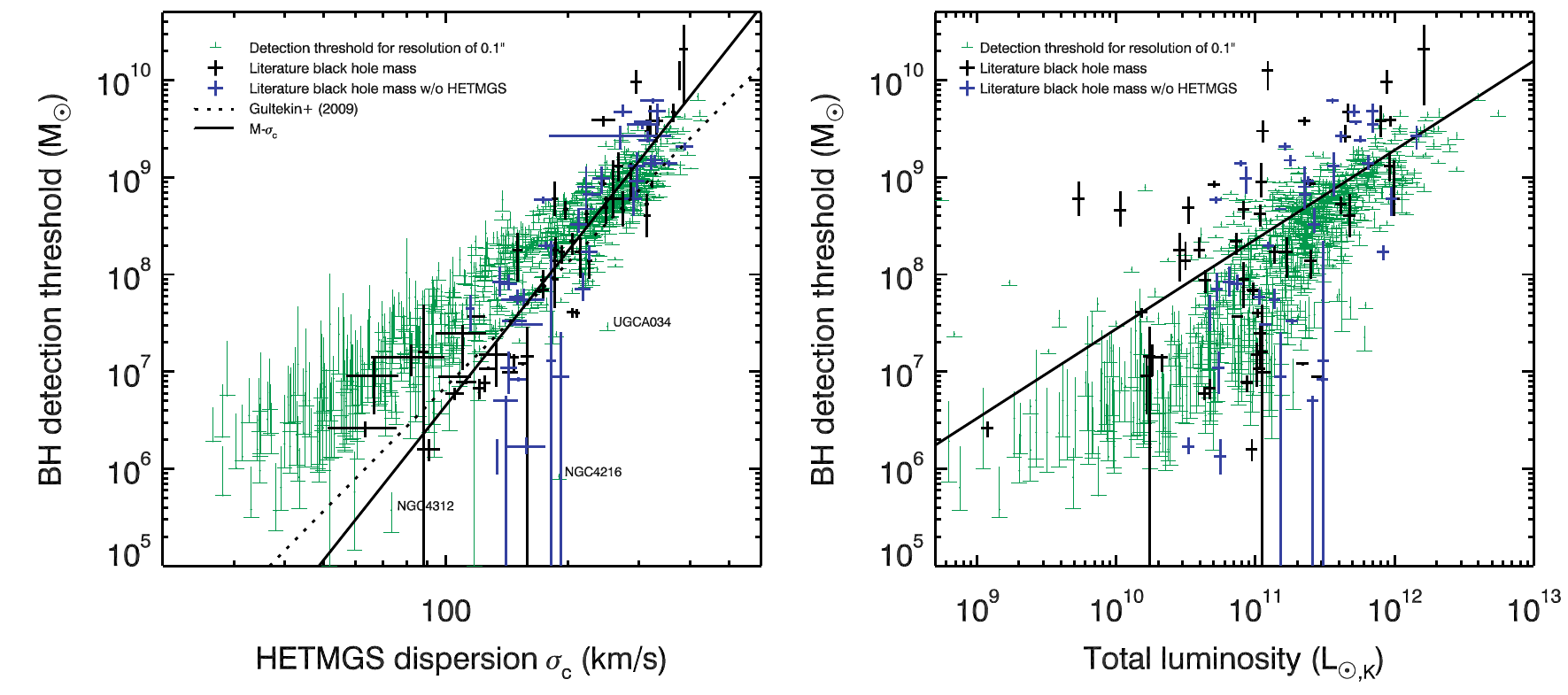}
\caption{The black hole mass detection threshold of the HETMGS galaxies in comparison to the \Msigma\ and \MLtot\ scaling relation {\protect \citep{2014ApJ...780...70L}}. This detection limit $M_t\equiv G^{-1} \theta_t D\sigma_c^{-2}$ assumes an angular resolution limit of a $\theta_t=0.1$\arcsec\ and the uncertainty in $M_t$ is derived from the uncertainty in $\sigma_c$. Known black hole masses are over-plotted. These thresholds are expected upper detection limits on possible black hole measurements: When the black hole is smaller than the threshold, its \SOI\ would not be resolved by current generation optical and infrared telescopes. It is apparent that very few low-dispersion galaxies exist that can constrain the \Msigma. There are also no large-dispersion galaxies that could leverage on \Msigma\ with an under-massive black hole. The right panel shows the same, but now for total luminosity. Here the situation is different and many objects can be used to discriminate different \MLtot\ relations. See \S\ref{sec:soi_distribution} for details.}
\label{fig:MLlimits}
\end{figure*}

\label{sec:black_hole_scaling_relations}

Galaxies with black hole measurements, hereafter referred to as host galaxies, have properties that are biased in comparison to the galaxy population as a whole. This is evident from the distribution of host galaxy properties versus the galaxy population, shown in figure~\ref{fig:Sample}. It is striking how the host galaxies trace out a very narrow locus in this parameter space. This is most obvious in the luminosity--size panel, where they lie along a narrow line, sampling preferentially the densest galaxies. Notice that the black hole host galaxies are typically denser than the average (early-type) galaxies. This severely hinders a robust measurement of the coefficients of higher dimensional scaling relations or non-linear scaling relations.

Target galaxies that can challenge the scaling relations are most interesting, especially if they can be shown to have black holes smaller than predicted from the scaling relations. For a given spatial resolution $\theta_t$, the black hole mass detection threshold $M_t$ can be defined as $M_t\equiv G^{-1} \theta_t D\sigma_c^{-2}$ by inverting the \SOI\ criterion (\S\ref{sec:the_soi_criterium}). This inversion is independent of any scaling relation and can thus be used to select galaxies that can challenge a particular scaling relation. Note that for a given host galaxy it is easier to detect an over-massive black hole, as its \SOI\ is exponentially easier to resolve.

In figure~\ref{fig:MLlimits} these lower limits are shown for \Msigma\ and \MLtot. It is notable that the detection threshold overlaps with the existing black hole measurements on \Msigma. This fact raises the question of whether or not the \Msigma\ relation can indeed be constrained at all. In particular, \cite{2010ApJ...711L.108B} posed that the \Msigma\ could just be an upper envelope relation. \cite{2011ApJ...738...17G} argued that such analysis does not take into account the relative scarcity of upper limits to black hole masses compared to the number of detections. Because the value of the black hole mass is unknown before measuring it, an upper envelope relation would predict many more upper limits than detections at a fixed velocity dispersion and distance; instead the opposite is found. When using information from upper limits and detected black hole masses, \citet{2011ApJ...738...17G} found the relation to be best described by a ridge-line relation and could rule out the upper envelope relation at $>99\%$ confidence. It is also clear that highly accurate black hole masses are needed to secure \Msigma\ at the low dispersion end. There is not a significant number of new galaxies in the survey with extremely large \SOIs\ to help discriminate between these two scenarios. There are no galaxies with a \SOI\ bigger than 0.1\arcsec\ and a dispersion below 90 \kms. Hence very few mass measurements exist for such low dispersion galaxies \citep[but see][]{2002ApJ...567..237S,2009ApJ...692..856B}.

The \MLtot\ and \MLrel\ relation have a similar issue at high luminosity. There are no high luminosity galaxies nearby enough for the detection of an under-massive black hole. In figure~\ref{fig:MLlimits} we show the lowest black hole masses that can be detected as a function of total luminosity. This is conservative, as only considering the bulge would decrease the luminosity, but does not change the black hole mass detection threshold\footnote{Bulge fractions are not known for the 2MASS and HETMGS galaxies, and thus we will only consider total luminosity here. Nonetheless, the \MLtot\ has been shown to be a good proxy for black hole mass by \cite{2014ApJ...780...70L}}. If luminous early types with small $10^6$ \Msun\ black holes exist, we would not be able to detect them \citep[e.g.][]{2011ApJ...741...38G}. There is a distinct lack of high luminosity galaxies in which a low mass black hole could be detected. This is because the closest large luminosity ellipticals are in Virgo at a distance of 14 Mpc, with the exception of Maffei I (UGCA034). The \MLtot\ predicts relatively massive black holes for the nearby spirals. These galaxies do not to host large bulges \citep{2010ApJ...723...54K} and are thus ideal to distinguish between \MLtot, \MLrel\ and \Msigma, as even an upper-limit would provide strongly leverage on \MLtot.

\section{Looking Ahead}
\label{sec:conclusions}

This paper introduces the HET Massive Galaxy Survey, which consists of long slit spectroscopy of \HETNGals\ nearby galaxies. The surveyed galaxies were specifically chosen for their potential for a direct dynamical black hole mass measurement. By selecting galaxies with large \SOIs , the survey provides the most complete prerequisite sample for future dynamical black hole mass measurements in the local volume. Other nearby galaxy surveys do not (specifically) probe galaxies that are nearby enough for black hole mass measurements and the HETMGS is thus complementary. The survey improves many velocity dispersion measurements present in the the HyperLeda catalog (\S\ref{sec:kinematics_comparison}), which is often used as the basis for black hole mass measurements. The sample is very diverse and spans a large range in luminosity, size, dispersion and galaxy morphology (\S\ref{sec:HETgalaxies}). The central stellar kinematics and emission line ratios are the survey's primary data products and are presented in table~\ref{tab:hetmgs}. The HET observations have been used in the following papers \cite{2012Natur.491..729V}, \cite{2013MNRAS.434L..31L}, \cite{WalshN1271} and \cite{Yildirim1}.

Currently there are only 90 galaxies with direct black hole masses measurements and the existing data do not favor a specific black hole scaling relation over another \citep{2013ARA&A..51..511K}. This in part a result of the fact that the host galaxies are heavily biased and sample only the densest of galaxies, which is not representative of the galaxy population at large (\S\ref{sec:black_hole_scaling_relations}). They are also strongly clustered in luminosity and size. Expanding black hole mass measurements to be more representative of the galaxy population, will greatly extend the leverage on the black hole scaling relations. This is currently an active field, with different teams pursuing different parts of parameter space \citep{2012ApJ...756..179M, 2010MNRAS.403..646N, 2013AJ....146...45R, 2009MNRAS.399.1839K, 2012Natur.491..729V, 2014Natur.513..398S, 2011ApJ...727...20K, WalshN1271}. Even with all the progress, the current crop of black hole masses is still confined to a narrow range of host galaxy properties. Unfortunately, the HETMGS survey shows that the potential targets are not very numerous. The black hole scaling relations are thus strongly limited by the detection thresholds set by angular resolution of telescopes. The increased spatial resolution offered by VLBI \citep{2011ApJ...727...20K}, ALMA \citep{2014arXiv1406.2555D} and the ELTs \citep{2014AJ....147...93D} are crucial to resolve the \SOI\ of low-mass black holes.

Also important are the the cross-calibrations between methods. These are crucial to independently verify the intrinsic uncertainties. Unfortunately, there are few objects where two or more of the main methods---gas dynamics, stellar dynamics, megamasers, and reverberation mapping---can be applied. The lack of such comparable measurements is a result of the different requirements that each method has concerning host galaxy properties. Comparisons between gas and stellar dynamical black hole masses have been done for only six objects \cite[and references therein]{2013ApJ...770...86W}. In half of those cases the gas dynamical masses are lower by at least a factor of two for reasons yet unknown.   So far only NGC\,4151 has both a dynamical and reverberation mass. However its stellar kinematics are strongly affected by the spiral perturbations and complex bar kinematics \citep{2007ApJ...670..105O, 2014ApJ...791...37O}. Comparison targets of megamasers galaxies are similarly rare \citep{2009ApJ...693..946S}. The HETMGS can be the starting point for finding new cross-calibration targets.

The first result of the HETMGS survey was the discovery of six extremely compact, high-dispersion, galaxies which are candidates to host black holes that are too large for their galaxy mass \citep{2012Natur.491..729V,2013MNRAS.431L..38F,2013MNRAS.433.1862E}. Apart from NGC1277, \cite{WalshN1271} found another compact galaxy with a over-massive black hole in the HETMGS sample. And there are hints for Mrk~1216 \citep{Yildirim1} and b19 \citep{2013MNRAS.434L..31L}.  As the HETMGS survey is extremely suitable to find the densest galaxies in the nearby volume, because it is heavily biased towards the densest systems.  These highly compact galaxies are very interesting, because they could be the passively evolved \referee{ancestors}\ of the quiescent galaxies at $z\sim2$ (red nuggets), sub-mm galaxies and quasars found at high redshift $z>4$ \citep{2012Natur.491..729V,2014ApJ...780L..20T, 2014ApJ...782...68T}. A detailed study of these curious objects is outside the scope of this paper. 

The HETMGS only contains Northern galaxies and an extension of the survey to the South would double the viable targets. This is very worthwhile given the rarity of suitable targets for dynamical black hole mass measurements and the availability of high resolution Sourthern facilities. In the South, galaxies with $m_{ks} < 11.7$ already have low resolution, single aperture, spectroscopy from the 6dF Galaxy Survey \citep{2009MNRAS.399..683J, 2014MNRAS.443.1231C} and the follow up can thus be planned efficiently.

In conclusion, this paper describes the HET Massive Galaxy Survey, including the sample selection, data reduction and derived quantities of \HETNGals\ galaxies. We also show that the current crop of black hole masses is strongly biased and that this survey is ideally suited to plan the next generation of direct black hole mass measurements.

\section*{Acknowledgments}

It is a pleasure to thank the Telescope Operators Frank Deglman, Vicki Riley, Eusebio Terrazas, Amy Westfall and Resident Astronomers, John Caldwell, Stephen Odewahn, Sergey Rostopchin,  Matthew Shetrone. We thank Arjen van der Wel, Ronald L\"asker and Glenn van de Ven for discussions and suggestions that improved the survey. 

The Hobby-Eberly Telescope (HET) is a joint project of the University of Texas at Austin, the Pennsylvania State University, Ludwig-Maximilians-Universit\"at M\"unchen, and Georg-August-Universit\"at G\"ottingen. The HET is named in honor of its principal benefactors, William P. Hobby and Robert E. Eberly. The Marcario Low Resolution Spectrograph is named for Mike Marcario of High Lonesome Optics who fabricated several optics for the instrument but died before its completion. The LRS is a joint project of the Hobby-Eberly Telescope partnership and the Instituto de Astronom\'ia de la Universidad Nacional Aut\'onoma de M\'exico. %\footnote{Based on observations obtained with the Hobby-Eberly Telescope, which is a joint project of the University of Texas at Austin, the Pennsylvania State University, Stanford University, Ludwig-Maximilians-Universit\"at M\"unchen, and Georg-August-Universit"at G\"ottingen.}

J.L.W. has been supported by an NSF Astronomy and Astrophysics Postdoctoral Fellowship under Award No. 1102845. This material is based upon work supported by the National Science Foundation under Grant No. AST-1107675.

The work greatly depended on the public databases, \href{http://leda.univ-lyon1.fr}{HyperLeda},  NASA's Astrophysics Data System and the NASA/IPAC Extragalactic Database (NED), which is operated by the Jet Propulsion Laboratory, California Institute of Technology, under contract with the National Aeronautics and Space Administration. This research has made use of NASA's Astrophysics Data System.
 
This publication makes use of data products from the Two Micron All Sky Survey, which is a joint project of the University of Massachusetts and the Infrared Processing and Analysis Center/California Institute of Technology, funded by the National Aeronautics and Space Administration and the National Science Foundation.

Funding for the \href{http://www.sdss.org/}{Sloan Digital Sky Survey} (SDSS) has been provided by the Alfred P. Sloan Foundation, the Participating Institutions, the National Aeronautics and Space Administration, the National Science Foundation, the U.S. Department of Energy, the Japanese Monbukagakusho, and the Max Planck Society. The SDSS is managed by the Astrophysical Research Consortium (ARC) for the Participating Institutions. The Participating Institutions are The University of Chicago, Fermilab, the Institute for Advanced Study, the Japan Participation Group, The Johns Hopkins University, Los Alamos National Laboratory, the Max-Planck-Institute for Astronomy (MPIA), the Max-Planck-Institute for Astrophysics (MPA), New Mexico State University, University of Pittsburgh, Princeton University, the United States Naval Observatory, and the University of Washington. %The SDSS Web site is \href{http://www.sdss.org/}.

\bibliography{smbh}

\begin{center}
{\scriptsize

\begin{longtable*}{llccrcccccc}  
\hline
\hline 
2MASS ID & Name & Frame & MJD & PA & slit & $V_{hel}$ & $\sigma_c$ & [NII]/H$\alpha$ & [OIII]/H$\beta$ & AoN \\
     &      &      & & (\dgr) & (\arcsec)& (\kms)    & (\kms)  & & & \\
 \hline                            
\endfirsthead
\hline 
2MASS ID & Name & Frame & MJD & PA & slit & $V_{hel}$ & $\sigma_c$ & [NII]/H$\alpha$ & [OIII]/H$\beta$ & AoN \\
     &      &      & & & (\arcsec) & (\kms)    & (\kms)  & & & \\
 \hline                            
\endhead
\hline \multicolumn{9}{r}{{Table continues on next page}} %\\ \hline
\endfoot  
\hline \hline
\endlastfoot
 
              \texttt{                  } & \texttt{             NGC5623 } & 050570 & 55309.407 &    197 &  2.0 &   3417.0$\pm$13.7 &    260.8$\pm$6.7 & 0.24$\pm$0.04 & 1.29$\pm$0.43 & 23,1 \\ %    168.7     -4.4   1.45
              \texttt{                  } & \texttt{             NGC5623 } & 050571 & 55309.418 &    197 &  2.0 &   3406.8$\pm$14.0 &    262.2$\pm$7.0 & 0.18$\pm$0.04 & 0.59$\pm$0.13 & 26,3 \\ %    167.8     -4.4   1.60
              \texttt{                  } & \texttt{             NGC5623 } & 051125 & 55327.364 &    197 &  2.0 &   3463.9$\pm$12.5 &    250.5$\pm$6.4 & 0.19$\pm$0.04 & 0.53$\pm$0.10 & 31,4 \\ %    170.1    -10.4   1.72
              \texttt{                  } & \texttt{             NGC5623 } & 051126 & 55327.375 &    197 &  2.0 &   3471.1$\pm$14.5 &    249.3$\pm$6.8 & 0.15$\pm$0.03 & 0.59$\pm$0.12 & 31,3 \\ %    169.5    -10.4   1.89
              \texttt{                  } & \texttt{            NGC5846A } & 341486 & 56329.472 &     90 &  1.0 &   2196.8$\pm$11.3 &    195.3$\pm$3.5 & -0.17$\pm$0.56 & 0.12$\pm$0.50 & 2,1 \\ %    106.9     28.6   1.96
              \texttt{    0223114+425931} & \texttt{            UGC01841 } & 060058 & 55413.410 &     90 &  2.0 &   6245.8$\pm$19.9 &    241.9$\pm$11.4 & 0.49$\pm$0.03 & 0.74$\pm$0.17 & 26,2 \\ %    180.8     26.0   1.80
              \texttt{    0223114+425931} & \texttt{            UGC01841 } & 060059 & 55413.422 &     90 &  2.0 &   6244.0$\pm$17.5 &    248.5$\pm$9.7 & 0.52$\pm$0.03 & 0.92$\pm$0.29 & 28,2 \\ %    181.2     26.0   1.49
              \texttt{    0223114+425931} & \texttt{            UGC01841 } & 060162 & 55418.384 &     90 &  2.0 &   6271.6$\pm$13.0 &    258.0$\pm$8.0 & 0.48$\pm$0.04 & 1.21$\pm$0.50 & 21,1 \\
\end{longtable*}
}
\end{center}

\end{document}